\documentstyle [a4,epsfig]{article}

\newcommand{\lab}{\label}

\newcommand{\bc}{\begin{center}}
\newcommand{\ec}{\end{center}}
\newcommand{\be}{\begin{equation}}
\newcommand{\ee}{\end{equation}}
\newcommand{\bea}{\begin{eqnarray}}
\newcommand{\eea}{\end{eqnarray}}
\newcommand{\bs}{\begin{subequations}}
\newcommand{\es}{\end{subequations}}
\newcommand{\beq}{\begin{eqalignno}}
\newcommand{\eeq}{\end{eqalignno}}

\def\al{\alpha}

\def\om{\omega}

\def\lab{\label}
\begin{document}

\thispagestyle{empty}

\vspace{2.0cm}
\bc

\huge{The dissipative quantum model of brain: how do memory localize 
in correlated neuronal domains}  

\vspace{1.2cm}

\large{ Eleonora Alfinito$^{1,2}$, and Giuseppe Vitiello$^{1,3}$} \\
\small
{\it ${}^{1}$Dipartimento di Fisica, Universit\`a di Salerno, 84100 
Salerno, Italy\\
${}^{2}$INFM Sezione di Salerno}\\
${}^{3}$INFN Gruppo Collegato di Salerno\\
alfinito@sa.infn.it\\
vitiello@sa.infn.it

\vspace{1.5cm}

\ec

\normalsize
{\bf Abstract}

The mechanism  of memory localization in extended domains is 
described in the framework of the parametric dissipative quantum 
model of brain. The size of the domains and the capability in 
memorizing depend on the number of links the system is able to 
establish with the external world.

\vspace{4.5cm}

\newpage

\section{Introduction}

It is experimentally well established, since Lashley and 
Pribram \cite{P2,PR} pioneering work, that many functional 
activities of the brain involve extended assembly of neurons. On this 
basis, Pribram has introduced concepts of Quantum Optics, such as 
holography, in brain modeling \cite{P2}. Information is indeed 
observed to be spatially uniform "in much the way that 
the information density is uniform in a hologram" \cite{fre1}. While 
the activity of the single neuron is experimentally observed in form 
of discrete and stochastic pulse trains and point processes, the 
``macroscopic'' activity of large assembly of neurons appears to be 
spatially coherent and highly structured in phase and amplitude 
\cite{fre2, fre3}. The quantum model of brain proposed in 1967 by  
Ricciardi and Umezawa \cite{UR} is firmly founded on such an 
experimental evidence. The model is in fact primarily aimed to the 
description of non-locality of brain functions, especially of memory 
storing and recalling. The mathematical 
formalism in which the model is formulated is the one of Quantum 
Field Theory (QFT) of many body systems. In one of his last papers 
\cite{UC} Umezawa explains the motivation for using QFT: "In any 
material in condensed matter physics any particular information is 
carried by certain ordered pattern   
maintained by certain long range correlation mediated by massless   
quanta. It looked to me that this is the only way to memorize   
some information; memory is a printed pattern of order supported   
by long range correlations...If I could know   
what kind of correlation, I would be able to write down the   
Hamiltonian, bringing the brain science to the level of condensed   
matter physics." The main ingredient of the model is thus the 
mechanism of spontaneous 
breakdown of symmetry by which long range correlations (the 
Nambu-Goldstone (NG) boson modes) are 
dynamically generated in many body physics. In the model the 
"dynamical variables" are identified \cite{YA} with
those of the electrical dipole vibrational 
field of the water 
molecules \cite{DG, PRL} and of 
other biomolecules present in the brain structures, and with the ones 
of the associated NG modes, named the dipole wave quanta (dwq) 
\cite{VT}. The model, further developed by Stuart, Takahashi and 
Umezawa \cite{S1,S2} 
(see also \cite{CH}), exhibits interesting features related with 
the r\^ole of microtubules in the brain 
activity \cite{P2,PR,YA} and its
extension to dissipative dynamics \cite{VT} allows a huge memory 
capacity. The dissipative 
quantum model of brain has been investigated \cite{PV} also in 
relation with the modeling of neural networks exhibiting long range 
correlations among the net units. One motivation for such a study is 
of course the great interest in neural network modeling, in
computational neuroscience and in quantum 
computational strategies based on quantum evolution (quantum 
computation)\cite{QC}. 

In the present paper, we consider the parametric extension of the 
dissipative quantum model of brain. In previous works it has been 
considered the case  of
time independent frequencies  associated to the dwq. A more general
case is the one of time-dependent frequencies. The dwq may in fact 
undergo a number of fluctuating
interactions and then their characteristic frequency
may  accordingly change in time. The aim of this paper is to show 
that dissipativity and frequency 
time-dependence lead to the dynamical organization of the memories in 
space (i.e. to their localization in more or less diffused regions of 
the brain) and in time (i.e. to their longer or shorter life-time). 

The results we obtain  agree with physiological observations 
which show that the neural connectivity is observed to grow as the 
brain develops and relates to the external world (see e.g. 
\cite{GRE}). In the non-parametric case the 
region involved in memory recording (and recalling) was extending to 
the full system. According to the results below presented, the 
non-locality of memory more realistically appears now in the 
dynamical formations of finite size correlated "domains". On the 
other hand, the finiteness 
of the size of these domains implies an effective non-zero mass for 
the dwq  and this in turn 
implies the appearance of related time-scales for the dwq propagation 
inside the domain. The arising of these time-scales seems to match 
physiological observation of time lapses observed in gradual 
recruitment of neurons in the establishment of brain functions 
\cite{LI, GRE}. Frequency time-dependence also 
introduces a fine structure in the decay behavior of memories, as we 
will see. A further result characteristic of our model is the one 
which shows that the psychological arrow of time points in the same 
direction in which point the thermodynamical arrow of time and the 
cosmological arrow of time (defined by the expanding 
Universe direction \cite{double, Hawking}).

Finally, we remark that some criticisms recently advanced 
\cite{teg} on the use of quantum formalism in brain modeling are 
quite easily turned down. Such criticisms are founded on the 
computation of the decoherence time of the neuron and of the 
microtubule. Such a decoherence time is found to be many order of 
magnitude 
shorter than typical dynamical times associated with neuron activity 
and kink-like microtubule excitations. The "conclusion" that neurons 
and microtubules are classical objects is "then" reached. As a matter 
of fact, Stuart, Takahashi and Umezawa have anticipated such a 
"discovery" noticing \cite{S1}, with a pleasant sense of humor, that 
"it is difficult to consider neurons as quantum objects". A careful 
reading of the literature thus shows that, since 1967 \cite{UR}, the 
conclusion of ref. \cite{teg} was taken to be a rather obvious fact 
by the authors of the papers where the quantum model of brain and its 
developments have been discussed.
The "quantum" variables entering the 
formalism are the dynamical variable mentioned above, not to be 
confused with neurons and other cells. The neurons are purposely not 
even considered to be "the fundamental units of the brain" \cite{UR}.

In the following Section, where we briefly discuss some aspects of 
the dissipative quantum model, we will further consider some of the 
motivations to use QFT in brain modeling. The parametric extension,  
the formation of domains and finite life-time modes are discussed in 
Section 3 and 4. Section 5 is devoted to final comments and 
conclusions.

\section{The dissipative quantum model of brain}

In the quantum model of brain \cite{UR} memory recording is
represented by the ordering induced in the ground 
state ("coding") by means of the
condensation of NG modes. These are dynamically generated through the 
breakdown of the rotational 
symmetry of the electrical dipoles of the water molecules and are called
dipole wave quanta (dwq). 
The trigger of the symmetry breakdown is the external 
informational input. The recall mechanism is described as the 
excitation of dwq from the ground state under the action of an 
external imput similar to the one which produced the memory 
recording.

The macroscopic behavior of the brain is thus derived from the 
microscopic dynamics of quantum fields. The "code" classifying the 
recorded information is identified with the "order parameter" which 
is the macroscopic variable defining the system (memory) state. 
The high stability of memory demands  
that the long range correlation modes (the dwq) must be in the 
lowest energy state (the  
ground state), which also guarantees that memory is easily  
created and readily excited in the recall process. 
The long range correlations must also be quite  
robust in order to survive against the state of continuous  
electrochemical excitation of the brain and the continual response 
to external stimulation. At the same time,  
however, such electrochemical activity must also, of course, be 
coupled to the  
correlation modes. It is indeed the electrochemical activity observed 
by neurophysiology that provides a first response to external 
stimuli. The brain is then modeled \cite{S1, S2} as a "mixed" system 
involving two separate but interacting levels. The memory level is a 
quantum dynamical level, the electrochemical activity is at a 
classical level. The interaction between the two dynamical levels is  
possible because the  
memory state is a {\it macroscopic quantum state} due to the {\it  
coherence} of the correlation modes.  

In many-body physics there are many systems  
whose macroscopic properties must be described classically, {\it but  
they can only be explained as arising from a quantum dynamics}. 
The crystals, the superconductors, the superfluids, the  
ferromagnets, and in general all systems presenting   
observable ordered patterns are systems of this kind. 
Of course, any physical system is, in a trivial sense, a quantum  
system since any system is made by atoms which are quantum  
objects. But it is not in this trivial sense that the  
above mentioned systems are macroscopic quantum systems. 
The specific, not at all trivial way in which  
they appear to be "macroscopic quantum systems" has to do with  
the {\it dynamical} origin of the macroscopic scale out of the  
microscopic quantum scale of the components. In other words, the 
macroscopic scale has to do not only with the large number of 
constituents assembled  
in the system (trivial summing up). This is a  
necessary, but not sufficient condition. The "emergence" of 
the macroscopic scale has to do  
{\it primarily} with the appearance of long  
range correlations among the microscopic constituents. Due to  
such correlations the rate of quantum fluctuations is negligible  
and the system behaves as a classic one. It is well known that long 
range correlation modes, and their stable coherent 
condensation in the lowest energy state, {\it cannot be understood  
without recourse to quantum dynamics}. 

It is then in such a way that the "classical" behavior of the  
memory state has to be understood. The 
density of the dwq condensed in the ground state represents the 
information code, namely the order parameter which is a macroscopic 
variable. The 
state appears {\it therefore} as a classically behaving macroscopic  
quantum state.  
 
The problem of the coupling between the quantum  
dynamical level and the electrochemical level is then  
reduced to the problem of the coupling between two macroscopic  
entities. Such a coupling is analogous, for example, to the coupling 
between  
classical acoustic waves and phonons in crystals. Acoustic waves  
are classical waves; phonons are quantum long range modes (the 
elastic wave quanta). Their coupling is very well known and of course 
experimentally observed \cite{wolfe}.

The interaction of the external stimulus with the brain is, in  
conclusion, mediated by the electrochemical response. This response 
sets the boundary conditions such that symmetry is broken  
firstly in limited regions (coherence domains). If enough energy 
comes into play, the  
coherence domain boundaries may be broken; the domains then  
merge into larger ordered regions with the establishment of  
long range correlation modes and consequent recording of the  
information.
 
We also remark that the quantum model of brain is a 
QFT model, {\it not} a Quantum Mechanics (QM) model. QFT is {\it 
dramatically} different from QM. In fact the von Neumann theorem 
states 
that all the representations of the canonical commutation relations 
are unitary equivalent (and therefore physically equivalent) in 
systems with a finite number of degrees of freedom, and therefore in 
QM. On the contrary, in QFT the number of degrees of freedom 
is infinite, the von Neumann theorem thus does not hold and there exist 
infinitely many unitarily inequivalent representations of the 
canonical commutation relations. It is because of the existence of 
the infinitely many 
unitarily inequivalent representations that in QFT a system may be in 
different physical phases , spontaneous symmetry breakdown can occur 
and dynamically generated ordering sustained by long range 
correlations may exist in a stable state. Only in QFT it is possible 
to describe "macroscopic quantum systems". These phenomena do not 
occur in QM. It is also known that the source of the 
non-perturbative nature of many phenomena resides in the 
manifold of inequivalent representations of QFT. The {\it 
decoherence} mechanisms studied in QM have thus no relation with the 
coherence mechanism studied in QFT. This is the founding 
basis of the QFT formalism used in the quantum brain model. 
Therefore, one can have ordered (memory) 
states, which are at the same 
time degenerate ground states, and thus stable states for the system. 
This last fact would be {\it in se} a strong motivation to use QFT in 
brain modeling (as in fact it was for Umezawa, cf. Section 1). It can 
be also shown \cite{PRL} that the time scale associated with the 
coherent interaction in the QFT of electrical dipole fields for water 
molecules is of the order of $10^{-14}$ $sec$, thus much shorter than 
times associated with short range interactions, and therefore these 
effects are well protected against thermal fluctuations. 

By taking into 
account the intrinsic dissipative character of the brain dynamics, 
namely that the brain is an {\it open system} 
continuously {\it linked} with (coupled to) the environment, the 
memory 
capacity can be shown \cite{VT} to be  enormously enlarged, 
thus solving one of the main 
problems 
left unsolved in the original formulation of the quantum brain model.
To see this let us denote the dwq variables by $A$ and recall that 
the canonical formalism for dissipative systems requires the 
introduction of a ``mirror'' set of 
dynamical variables, say $\tilde A$. The number ${\cal{N}}_{A}$ of 
$A$-modes, condensed in the vacuum,
constitutes the ``code'' of the information. 
The vacuum state is defined to be the state in which the 
{\em difference} ${\cal{N}}_{A} -{\cal{N}}_{\tilde A}$ is zero.   
There are thus infinitely many ground states, each one corresponding 
to a different value of the code ${\cal{N}}_{A}$.
The "brain  (ground) state" may be represented as
the collection (or the  superposition) of the full set of memory 
states ${|0>}_{\cal N}$, for all  $\cal N$.
 
The brain is thus described as a complex system with a 
huge number of macroscopic states (the memory states). The degeneracy
among the vacua ${|0>}_{\cal N}$ plays a crucial r\^ole in solving
the problem of memory capacity.  The  dissipative dynamics introduces
$\cal N$-coded "replicas" of the system   and information printing 
can be  performed in each replica without destructive interference 
with previously recorded informations in other replicas. A huge 
memory capacity is thus achieved \cite{VT}. 

As we will see the parametric extension of the dissipative quantum 
model leads to the formation of correlated domains of finite size, 
and to a fine structure in the life-time of the $A$ modes.

\section{The parametric extension of the dissipative model and the 
Bessel equations}

As mentioned above, in the dissipative model the canonical formalism 
requires the doubling of the system degrees of freedom \cite{VT}. 
Thus we are led to consider a couple of damped harmonic oscillators 
describing the system variable and its doubled, or mirror image. In 
the parametric model the associate frequency is assumed to be 
time-dependent \cite{brain2000}. The equations are then: 
\bea
\stackrel{..}u\,+\,L \stackrel{.}u\,+\, {\omega_{n}}^{2}(t) u \,=
&\ 0,
\nonumber \\
\stackrel{..}v\,-\,L \stackrel{.}v\,+\, {\omega_{n}}^{2}(t) v \,=
&\ 0,
\lab{qm3}\eea 
where:
\be 
  \omega_{n}(t)\ =\ {\omega}_{0} \ e^{-\frac{L 
t}{2n+1}}.
\lab{qm4} \ee
The quantities $u, v$ and $\omega_{0}$ are considered for fixed 
momentum $k$. $u$ and $v$ are related with the $A$ and $\tilde A$ 
modes when quantization is performed. In the following we will 
comment on the assumed exponential time-dependence of the frequency 
$\omega_{n}(t)$. Note that $\omega_{n}(t)$ approaches to the 
time-independent 
value $\om_{0}$ for $n \rightarrow \infty$: the 
frequency time-dependence is thus "graded" by $n$. We will see that 
$n$ has an interesting physical interpretation. $L$  
is a characteristic 
parameters of the system. Remarkably, it is found that the couple of 
equations (\ref{qm3}) is equivalent to the 
the spherical Bessel equation of order $n$ ($n$ is integer or zero): 
\be
z^2 \ {d^{2} \over {dz}^{2}}M_{n} + 2z \ {d \over {dz}}M_{n} + [z^2  
- n(n+1)] 
\ 
M_{n} \ = \ 0 ~.
\lab{qm42}\ee
As it is well known, Eq. (\ref{qm42}) 
admits as solutions a complete 
set of (parametric) decaying functions \cite{Abram,Jackson};
particular solutions are the first and second kind Bessel 
functions, or their linear combinations (the Hankel functions).

Note that both $M_{n}$ and $M_{-(n+1)}$ are solutions of the same eq. 
(\ref{qm42}). 
By using the substitutions: $w_{n,l}= M_{n}\ \cdot
(x_{n})^{-l}$, $z =\epsilon_{n} x_{n}$ and $x_{n} =  e^{-t
/\al_{n}}$, 
where $\epsilon_{n}, \al_n$ are arbitrary 
parameters, eq.(\ref{qm42}) goes into the 
following one:
\be
\stackrel{..}w_{n,l} \ - \ \frac{2l+1}{\al_{n}}\ \stackrel{.} w_{n,l} 
+ 
\left[\frac{l(l+1)-n(n+1)}{\al^{2}_{n}} \ +\   ( 
\frac{\epsilon_{n}}{\al_{n}})^{2}\ e^{-2t/\al_{n}} \right] \ w_{n,l}\ 
= \ 0,
\lab{qm7}\ee
where $\stackrel{.} w$ denotes derivative of $w$ with respect to 
time. 
Making the choice $l(l+1)=n(n+1)$ the degeneracy between the 
solutions 
$M_{n}$ and $M_{-(n+1)}$ is removed and two different equations are 
obtained,
one for $l=n$ and the other one for $l=-(n+1)$ ($l$ plays the r\^ole 
of a ``mirror'' index): 
\bea
&\stackrel{..}{ w}_{n,-(n+1)} \ + \ \frac{2n+1}{\al_{n}}\ 
\stackrel{.}{ w}_{n,-(n+1)} + 
[( \frac{\epsilon_{n}}{\al_{n}})^{2}\ e^{-2t/\al_{n}} ] \ w_{n,-
(n+1)}\ = 
\ 0,
\nonumber\\
&\stackrel{..}{ w}_{n,n} \ - \ \frac{2n+1}{\al_{n}} \ 
\stackrel{.}{w}_{n,n} + 
[( \frac{\epsilon_{n}}{\al_{n}})^{2}\ e^{-2t/\al_{n}} ] \ w_{n,n}\ = 
\ 0 
~.
\lab{qm6}\eea
By setting $u \equiv { w}_{n,-(n+1)},$ and $v \equiv { w}_{n,n}$, and 
choosing
the arbitrary parameters $\al_{n}$ and $\epsilon_{n}$ in such a way 
that 
$\frac{2n+1}{\al_{n}} \equiv L$ and 
$\frac{\epsilon_{n}}{\al_{n}} \equiv \om_0$ do not depend on $n$ 
(and on time), we see that 
eq. (\ref{qm6}) are nothing else than the couple of eqs. (\ref{qm3}) 
of 
the 
dissipative model with time-dependent frequency 
$\omega_{n}(t)$.

We note that $\al_{-(n+1)}=-\al_{n}$ and that the transformation 
$n \rightarrow -(n+1)$ leads to solutions (corresponding to $M_{-
(n+1)}$) 
which we will not consider since they have frequencies which are 
exponentially increasing in time (cf. eq. (\ref{qm4})). These 
solutions can be 
respectively obtained from the ones of eq. (\ref{qm3}) by 
time-reversal $t \rightarrow -t$.

We finally note that the functions $w_{n,n}$  and $w_{n,-
(n+1)}$ are "harmonically
conjugate" functions in the sense that they may be represented as 
$w_{n,-(n+1)}(t) = 
{1\over {\sqrt 2}} r_{n}(t)e^{{-L t\over 2}}$, \quad
 $w_{n,n}(t) = {1\over {\sqrt 2}} r_{n}(t)e^{{L t\over2}}$, 
respectively, with ${r}_{n}$ satisfying the parametric oscillator 
equation
\be
\stackrel{..}{r}_{n} +\Omega_{n} ^2(t)r_{n}=  0 ~,
\lab{qmm5}
\ee 
\be 
  \Omega_{n}(t) = \left [ \left ({\omega}_{n}^{2}(t)  - {{L  
  ^{2}}\over{4}} \right ) \right ]^{1\over{2}}.
  \lab{qm5} \ee

The quantization of the system (\ref{qm3}) can be now performed along 
the same line presented in refs. (\cite{double,brain2000,QD,PLA}).
The main feature of the dissipative quantization is the ``foliation'' 
of the representations \cite{QD}. Namely, at each time $t$ the system 
ground state is labeled by $t$, $|0(t)>$, so that at $t' 
\neq t$ the ground state $|0(t')>$ is unitary inequivalent to 
$|0(t)>$: in its time evolution the system runs over unitarily 
inequivalent representations. The generator of such a non-unitary 
time evolution is found to be related to the entropy variation rate, 
as it should be expected since dissipation implies irreversibility 
\cite{QD}. We thus see that the {\it arrow of time} naturally emerges 
in the dissipative quantum model. Moreover, it can be shown that the 
system ground state is also a thermal state \cite{QD, VT}, and the 
arrow of time due to dissipation is actually concord with the 
thermodynamical arrow of time (pointing in the increasing entropy 
direction). It has been shown \cite{double} that both these arrows 
point in the same direction of the cosmological arrow of time.

We remark that the dissipative time evolution cannot be described in 
the framework of Quantum Mechanics, since there all the 
representations are unitarily equivalent due to the von Neumann 
theorem (cf. Section 2). Thus the motivation to use QFT in brain 
modeling is reinforced.

In the infinite volume limit, due to the representation unitary 
inequivalence, any transition among degenerate vacua would be 
strictly forbidden. 
However, in realistic conditions  non-unitary time evolution and 
realistic phase transitions are possible due to boundary effects 
which "smooth out" the infinite volume limit and inequivalence among 
representations is accordingly also smoothed out. We thus observe the 
appearance of boundaries, namely of finite size domains.
This is discussed in some more details in the next section.

Finally, let us observe that 
the time-dependence of the frequency $\Omega_n$ of the 
coupled systems means that 
energy is not conserved in time and therefore that the $A-{\tilde A}$ 
system does not constitute a "closed" system. However, when $n 
\rightarrow \infty$, $\Omega_n$ approaches to a time independent 
quantity, which means that energy is conserved in such a limit, i.e. 
the $A-{\tilde A}$ system gets "closed" in that limit. Thus, in the 
limit $n \rightarrow \infty$ the possibilities of the system $A$ to 
couple to ${\tilde A}$ (the environment) are "saturated": the system 
$A$ gets {\it fully} coupled to $\tilde A$. This suggests that $n$ 
represents the number of {\it links} between $A$ and ${\tilde A}$. 
When $n$ is not very large (infinity), the system $A$ (the brain) has 
not fulfilled its capability to establish links with the external 
world (represented by $\tilde A$). On the other hand, as already 
mentioned, from eq.(\ref{qm4}) 
we also see that $n$ ``graduates'' the exponential dependence on time 
of the frequency $\omega_n(t)$, namely the rate of variations in time 
of the frequency, or, equivalently, the "rapidity" of the system 
response to  external stimuli. This has been the reason for our 
assumption of the exponential dependence on time of the frequency 
$\omega_n(t)$.

\subsection{Domains and Life-time}

We observe that in order the memory recording may occur, the 
frequency (\ref{qm5}) has to be real. Such a reality 
condition is found to be satisfied only in a definite span of
time, i.e., upon restoring the suffix $k$, for times $t$ such that 
$0 \le  t \le T_{k,n}$, with $ T_{k,n}$ 
given by
\be
T_{k,n}\,=\,\frac{2n+1}{L}\ \ln\left(\frac{2\om_{0,k}}{L}\right). 
\lab{qm8}\ee
Thus, the  memory recording processes can occur in limited time 
intervals which have  
$T_{k,n}$ as the upper bound, for each $k$. For times greater than 
$T_{k,n}$ memory recording cannot occur. Note that, for fixed $k$, 
$T_{k,n}$ grows linearly in $n$, which means that the time span 
useful for memory recording (the ability of memory 
storing) grows  as the number of links which 
the system is able to lace with the external world grows: more the 
system is "open" to the external world (more are the links), better 
it can memorize (high ability of learning).

We can also see that a threshold exists for 
the $k$ modes of the memory process. In fact, the reality condition 
${\Omega_{k,n}}^{2}(t)
\ge 0$ implies that $k \ge {\tilde k}(n,t) \equiv k_{0} 
e^{{{L \over {2n+1}}t}}$, with $ k_{0} \equiv {L \over  {2c}}$ at any 
given $t$ (note that $\omega_{0}= kc$). 
Note that such a kind of "sensibility" to external stimuli only 
depends on the internal parameter $L$.
 
We remark that this intrinsic infrared cut-off precludes infinitely 
long wave-lengths (infinite volume limit). In fact only wave-length 
smaller than or equal to the cut-off ${\tilde \lambda} \propto 
{1\over {{\tilde k}(n,t)}}$ are allowed. This means that (coherent) 
domains of sizes less 
or equal to ${\tilde \lambda}$ are involved in the memory 
recording, and that such a cut-off shrinks in time for 
a given $n$. On the other hand, a growth of $n$ opposes to such a 
shrinking. These cut-off changes correspondingly reflect on the 
memory domain sizes. This also implies that transitions through 
different vacuum states (which would be unitarily inequivalent vacua 
in the infinite volume limit) at given $t$'s 
become possible. As a consequence, both the phenomena of association 
of memories and of confusion of memories, which would be avoided in 
the regime of strict unitary inequivalence among vacua (in the 
infinitely long wave-length regime), become possible \cite{VT}.

  \begin{figure}[t]
 \caption{``Lives'' of $k$ modes, for fixed $n$}
\epsfig{file=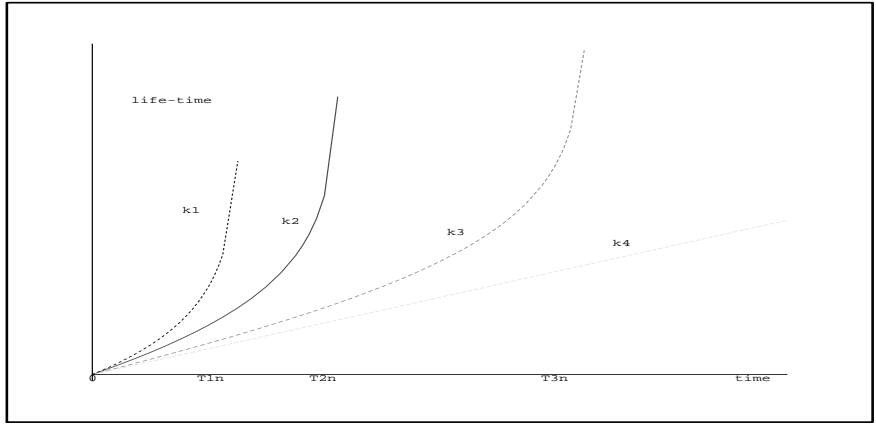,width=0.5\linewidth,height=1.0\linewidth,
angle=-90}  
  \end{figure}

We can estimate the domain evolution by introducing the quantity 
$\Lambda_{k,n}(t)$\cite{PLA, double}: 
  \be 
   e^{-2\Lambda_{k,n}(t)} \ =\  
  \frac{e^{-t \frac{L}{2n+1}}\,
{\rm sinh}\frac{L}{2n+1}({\rm T}_{k,n}-t)}{{\rm sinh} 
  \frac{L}{2n+1}{\rm T}_{k,n}}, \qquad \, \Lambda_{k,n}(t) \ 
\ge 0, \,{\rm   for \; any \;~ t} ~,
  \lab{c6}\ee 
i.e. $\Lambda_{k,n}(0)=0$ for any $k$, and 
$\Lambda_{k,n}(t)\rightarrow\infty$ for
$t \rightarrow\ T_{k,n}$ for any given $n$.
Then $\Omega_{k,n}$ may 
be expressed in the form: 
\be {\Omega_{k,n}(\Lambda_{k,n}(t))}\,=\,  \Omega_{k,n}(0) 
e^{-\Lambda_{k,n}(t)}~. 
  \lab{c5}\ee 
\begin{figure}[t]
  \caption{``Lives'' of $k$ modes, for growing $n$ and fixed $k$}
  
\epsfig{file=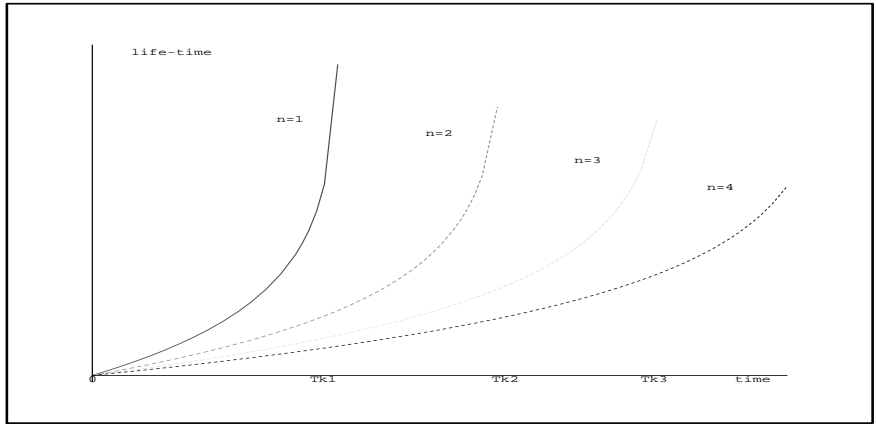,width=0.5\linewidth,height=1.0\linewidth,
angle=-90}
\end{figure}
%
%
%

Since ${\Omega_{k,n}(\Lambda_{k}({\rm T}_{k,n}))}\,=\,0$, we 
see that $\Lambda_{k,n}(t)$ acts as a life-time, say $\tau_{k,n}$, 
with 
$\Lambda_{k,n}(t) \propto \tau_{k,n}$,
for the mode $k$.
Modes with larger $k$ have a "longer" life with reference 
to time $t$. 
In other words, each $k$ mode "lives" with a proper time 
$\tau_{k,n}$, so 
that the mode is born when $\tau_{k,n}$ is zero and it dies for 
$\tau_{k,n}$ 
$\rightarrow$ $\infty$. 

The "lives" of the $k$ modes are sketched in the figures 
1-2. In Fig.1 the lives are drawn for growing $k$ and fixed $n$; 
vice-versa, in Fig.2, they are drawn for growing $n$ and fixed 
$k$.
The $\Lambda_{k,n}$s are drawn versus $t$, 
reaching the blowing up values in correspondence of the abscissa 
points 
$T_{k,n}$. Only the modes satisfying the reality condition are 
present at certain 
time 
$t$, being the other ones decayed.
This introduces an hierarchical organization of memories 
depending on their life-time: memories with a specific spectrum of 
$k$ 
mode components 
may coexist, some of them "dying" before, some other ones persisting 
longer. As observed above, since 
smaller or larger $k$ modes 
correspond to larger or smaller waves lengths, respectively, 
the (coherent) associated  memory domain sizes 
are correspondingly larger or smaller.

\section{Final remarks and conclusions}

In our model the processes of learning are each other independent, so 
it is possible that the ability in information recording may 
be different under different circumstances, at different ages, and so 
on. An interesting feature of the model which emerges from our 
discussion is that a higher or lower {\it 
degree of openness} (measured by $n$) to the external world may 
indeed produce a better or worse ability in learning, respectively 
(e.g. during the childhood or the older ages, respectively). 
We have also seen that the memory non-locality is "graded" by $n$ and 
by the spectrum of their $k$ modes components. These also control the 
memory life-time or persistence: more persistent memories 
(with a spectrum more populated by the higher $k$ components) are 
also more "localized" than shorter term 
memories (with a spectrum more populated by the smaller $k$ 
components), which instead extend over larger domains. 
It is thus expected that, for given $n$, "more impressive" is 
the external stimulus (i.e. stronger is the coupling with the 
external world) 
greater is the number of high $k$ momentum excitations produced in 
the brain and more "focused" is the "locus" of the memory.

The qualitative behaviors and results above 
presented appear to fit well with the physiological 
observations \cite{GRE} of the formation of 
connections among neurons as a consequence of the establishment of 
the links between the brain and the external world. More the brain 
relates to external environment, more neuronal connections will form. 
Connections appear thus more important in the brain functional 
development than the 
single neuron activity. Here we are referring to 
functional or effective connectivity, as opposed to the structural or 
anatomical one \cite{GRE}. In fact, while the last one can be 
described as quasi-stationary, the 
former one is highly dynamic with modulation time-scales 
in the range of hundreds of milliseconds \cite{GRE}. 
Once these functional 
connections are formed, they are not necessarily fixed. On the 
contrary, they may quickly change in a short time and new 
configurations of connections may be formed extending over a domain 
including a larger or a 
smaller number of neurons. Such a picture finds a possible 
description in our model, where the coherent domain formation, size 
and life-time depend on 
the number of links that the brain sets with its environment and  
on internal parameters.

The finiteness of the domain size also implies a non-zero effective 
mass of the dwq. These therefore propagate through the domain 
with a greater ``inertia'' than 
in the case of infinite volume where they are massless. The domain 
correlations are consequently established with a certain time-delay. 
This is also in agreement 
with physiological observations showing that the recruitment of 
neurons in a correlated assembly is achieved with a certain delay 
after the external stimulus 
action \cite{LI, GRE}. In connection with the recall mechanism, we 
note (see ref. \cite{VT}) that the dwq effective non-zero mass acts 
as a threshold in the excitation energy of dwq so that, in order to 
trigger the recall process  
an energy supply equal or greater than such a
threshold is required. When the energy supply is lower than the 
required threshold a "difficulty in recalling" may be experienced.  
At the same
time, however, the threshold may positively act as a "protection"
against unwanted perturbations (including thermalization) and
cooperate to the stability of the memory  state.  In the case of zero
threshold (infinite size domain) any replication signal could excite  
the recalling  and the brain would fall in a state of "continuous 
flow of memories". 

Finally, we note that {\it after} information has been 
recorded, the brain state is completely
determined and the brain cannot be brought to the state 
in which it was {\it before} the information printing occurred.  
Thus, one is actually obliged to consider the dissipative,  
irreversible time-evolution: the same fact of getting information 
introduces {\it the arrow of time} into brain dynamics. In other 
words, it introduces a partition in the time evolution, 
namely the {\it distinction} between the past and the future, a
distinction which did not exist {\it before} the information 
recording.
It can be shown that dissipation and the frequency 
time-dependence imply 
that the evolution of the memory state is controlled
by the entropy variations \cite{VT}: this feature reflects indeed the
irreversibility of time evolution (breakdown of time-reversal 
symmetry). 
The stationary condition for the free energy functional leads 
\cite{VT} then to recognize the memory state  $|0(t)>_{\cal N}$ to be 
a finite temperature state \cite{U}, which opens the way to the 
possibility of thermodynamic 
considerations in the brain activity.
In this connection we observe that the ``psychological arrow of 
time'', naturally emerging in the brain dynamics, turns out to point 
in the same direction of the ``thermodynamical arrow of time'', 
which points in the increasing entropy 
direction, and of the "cosmological arrow of time", defined by the 
expanding 
Universe direction \cite{Hawking}.

\end{document}